\documentclass[superscriptaddress,aps,prl,reprint,twocolumn,amsmath,amssymb]{revtex4-2}
\usepackage{graphicx}
\usepackage{hyperref}
\usepackage{amssymb}
\usepackage{slashed}
\usepackage{dcolumn}
\usepackage{amsmath}
\usepackage{bm}
\usepackage{colordvi}
\usepackage{algorithm}
\usepackage{algpseudocode}
\usepackage{multirow}
\usepackage{titlesec}
\usepackage{newtxtext,newtxmath}
\usepackage{dsfont}
\usepackage{xcolor}
\usepackage{mathbbol}

\usepackage{hyperref}
\hypersetup{
    colorlinks=true,
    linkcolor=blue,
    citecolor=blue,     
    urlcolor=blue,
}

\allowdisplaybreaks

\usepackage{mathrsfs}
\makeatletter

\newcommand{\Rmnum}[1]{\expandafter\@slowromancap\romannumeral #1@}
\makeatother

\begin{document}

\title{Polar nanoregions and reentrant-like ferroelectric behavior in  SrTiO$_3$}

\author{Yuan-Jie Sun}
\affiliation{Department of Materials Science and Engineering and Materials Research Institute, The Pennsylvania State University, University Park, PA 16802, USA}

\author{Fei Yang}
\email{fzy5099@psu.edu}
\affiliation{Department of Materials Science and Engineering and Materials Research Institute, The Pennsylvania State University, University Park, PA 16802, USA}

\author{Long-Qing Chen}
\email{lqc3@psu.edu}

\affiliation{Department of Materials Science and Engineering and Materials Research Institute, The Pennsylvania State University, University Park, PA 16802, USA}

\date{\today}

\begin{abstract}
Recent real-space imaging in quantum paraelectric SrTiO$_3$ [Nature {\bf 656}, 54 (2026)] reveals that local polar textures do not continuously grow upon cooling, but reach a maximum intensity at intermediate temperatures around 60-65~K and weaken again toward the quantum paraelectric ground state.   Such a reentrant-like  weakening
of local polar textures challenges the conventional paradigm in which ordering tendencies generally continue to strengthen as temperature decreases. Here we employ a self-consistent phase-field  theory showing that this anomalous behavior naturally arises from the coupling between intrinsic polar and antiferrodistortiv (AFD) fluctuations. We demonstrate that the flexoelectric-like coupling strongly hybridizes the polar and AFD modes. As the uncoupled polar and AFD modes cross near 52~K, their hybridization is maximized, driving the lower hybridized branch to develop a minimum on a finite-wave-vector shell. This finite-$q$ softening triggers a Brazovskii-type instability, strongly enhancing polarization correlations and producing nanoscale polar textures. Away from the crossing temperature, the two modes become increasingly detuned, weakening their hybridization and the associated finite-$q$ softening.  These results reaveal the origin of the formation of polar nanoregions in SrTiO$_3$, naturally explaining the unexpected confinement to an intermediate-temperature window and providing a mechanism beyond the quantum-fluctuation-based interpretation suggested by  experiment. Furthermore, we predict an unconventional reentrant-like sequence in weakly strained SrTiO$_3$, evolving from ferroelectric to paraelectric, polar-nanoregion, and eventually paraelectric regimes upon heating from zero temperature.
\end{abstract}

\maketitle

{\sl Introduction.---}Cooling a material toward its ground state generally enhances ordering tendencies by suppressing the related thermal fluctuations. Consequently, order parameters and correlation lengths typically increase monotonically upon cooling. Recent real-space imaging of SrTiO$_3$, however, has revealed a striking violation of this conventional expectation. As a prototypical quantum paraelectric, SrTiO$_3$ lies close to a ferroelectric quantum critical point~\cite{muller1979srtio3,yamada1969neutron,rowley2014ferroelectric,zhong1996effect,yang2026self,salje2013domains,barrett1952dielectric}, where the transverse optical (TO) soft mode continuously softens upon cooling but remains uncondensed in the low-temperature limit. Surprisingly, the experiments realized that polar nanoregions (PNRs) emerge below approximately 105~K, reach maximum strength at intermediate temperatures of 60-65~K, and are subsequently suppressed upon further cooling toward the quantum-paraelectric ground state~\cite{zhang2026imaging}. This reentrant-like weakening of local polar textures presents a fundamental challenge: why do polar correlations emerge only within an intermediate temperature regime, yet disappear at low temperatures, where thermal fluctuations are suppressed and ordering tendencies would normally be expected to strengthen? 

This anomaly has been attributed by experiments to the increasing  quantum fluctuations (zero-point fluctuations) upon approaching zero temperature~\cite{zhang2026imaging}, which are proposed to suppress local polar order and drive the disappearance of polar nanoregions. However, such an interpretation is unlikely, as an increasing accumulation of fluctuations upon cooling violates the fundamental thermodynamic principle of entropy reduction, irrespective of whether these fluctuations are described by classical or quantum statistics~\cite{abrikosov2012methods}. In both quantum field theory~\cite{peskin2018introduction} and condensed matter physics~\cite{abrikosov2012methods,mahan2013many}, zero-point fluctuations are intrinsic to the quantum ground state. Once this ground state is established, thermodynamic properties and thermal fluctuations arise from excitations above it, while the ground state itself already incorporates the renormalization effects of zero-point fluctuations~\cite{verdi2023quantum,wu2022large,yang2026self,yang2025efficient,yang2025microscopic,yang2025preformed,yang2025tractable}.

Theoretically, a natural mechanism for nanoscale polar textures is provided by finite-wave-vector instabilities. Unlike conventional ferroelectric ordering driven by a zone-center soft mode~\cite{cochran1981soft,cochran1961crystal,cowley1996phase,cowley1965theory,cochran1969dynamical,cochran1960crystal}, a softening at finite momentum can generate spatially modulated correlations and nanoscale structures. This is reminiscent of the Brazovskii framework~\cite{brazovskii1975phase,brazovskii1975phaset}, in which soft modes accumulate on a momentum shell satisfying $|\mathbf q|=q_c$, enhancing fluctuations and promoting modulated phases~\cite{fredrickson1987fluctuation,janoschek2013fluctuation,gruner1988dynamics,gruner1994dynamics}. Recent  studies~\cite{axe1970anomalous,zubko2007strain,yudin2013fundamentals,guzman2023lamellar} have proposed that flexoelectric coupling between the nearly critical ferroelectric soft mode and acoustic phonons may induce such finite-$q$ softening in SrTiO$_3$, potentially leading to a Brazovskii-type instability and modulated polar states. Experimental observations of local acoustic softening at finite momentum further support this possibility ~\cite{fauque2022mesoscopic,orenstein2025observation}. However, these mechanisms still predict, as thermodynamically expected, a continuously increasing tendency toward finite-$q$ polar ordering upon cooling, and therefore cannot account for the disappearance of PNRs in the low-temperature quantum-paraelectric regime.

Here we show that the essential ingredient is the interplay between ferroelectric and antiferrodistortive (AFD) lattice instabilities. Specifically, SrTiO$_3$ undergoes an AFD transition near $T_{\rm AFD}\approx105$~K~\cite{lytle1964x,shirane1969lattice,sai2000first,vogt1995refined,cowley1996phase}, driven by rotations of TiO$_6$ octahedra and a zone-boundary phonon instability at the $R$ point of the Brillouin zone. In particular, below $T_{\rm AFD}$, the AFD mode gradually hardens upon cooling, while the ferroelectric soft mode continues to soften. Such opposite temperature evolution of the two lattice instabilities suggests that their interplay may play a crucial role in shaping the unusual evolution of polar correlations in SrTiO$_3$. Accordingly, we employ a self-consistent phase-field theory incorporating coupled ferroelectric and AFD fluctuations. As the dispersions of two competing lattice modes intersect, flexoelectric-like coupling repels the two branches, driving the lower branch toward a finite-$q$ instability with a nearly degenerate shell of soft modes, a hallmark of Brazovskii physics~\cite{brazovskii1975phase}. Consequently, upon cooling from high temperatures, nanoscale polar textures emerge only within a finite intermediate temperature window where the two modes cross and strongly hybridize. Upon further cooling, the modes move away from resonance, the hybridization weakens, and the finite-$q$ instability collapses, naturally explaining the otherwise counterintuitive disappearance of polar nanoregions at low temperatures. Moreover, we also predict a reentrant-like sequence in weakly strained SrTiO$_3$, evolving from ferroelectric to paraelectric,
polar-nanoregion, and eventually paraelectric regimes upon heating from $T=0$.  These results reveal a distinct mechanism by which interactions between collective modes can create and subsequently extinguish local order upon cooling, providing an unexpected route to reentrant nanoscale textures in quantum materials. 

{\sl Model.---}We consider the polarization ${\bf P}({\bf r},t)$ and AFD distortion ${\bf Q}({\bf r},t)$ as the relevant lattice degrees of freedom. Their effective potential Hamiltonian is written as
\begin{equation}
\mathcal{H}
=\int\,d^3r
\left[\mathcal{V}_P({\bf P})
+\mathcal{V}_Q({\bf Q})
+\mathcal{V}_{\rm int}({\bf P},{\bf Q})
\right],
\end{equation}
The effective potential densities associated with the polarization and AFD distortions are given by~\cite{yang2026self,rowley2014ferroelectric}
\begin{eqnarray}
\mathcal{V}_P
&=&
\frac{a_P(T)}{2}|{\bf P}|^2
+
\frac{u_P}{4}|{\bf P}|^4
+
\frac{G_P}{2}
(\partial_iP_j)(\partial_iP_j),
\\
\mathcal{V}_Q
&=&
\frac{a_Q(T)}{2}|{\bf Q}|^2
+
\frac{u_Q}{4}|{\bf Q}|^4
+
\frac{G_Q}{2}
(\partial_iQ_j)(\partial_iQ_j),
\end{eqnarray}
where $a_{P,Q}(T)$ are the quadratic coefficients, $u_{P,Q}$ are the quartic coefficients, and $G_{P,Q}$ are the gradient stiffnesses. In the normal-mode basis, the harmonic Hamiltonian is diagonal, and the leading symmetry-allowed coupling between the transverse polar and AFD modes is a gradient-assisted three-phonon interaction in the centrosymmetric parent phase,  
\begin{equation}
\mathcal{V}_{\mathrm{int}}
=
\lambda_0\,{\bf P}\cdot\left[{\bf Q}\times\left(\nabla\times{\bf Q}\right)\right]+\lambda_1\,{\bf Q}\cdot\left[{\bf P}\times\left(\nabla\times{\bf P}\right)\right],
\end{equation}
where $\lambda_{0,1}$ characterizes the coupling strength. Expand the order parameters about their uniform configurations,
\begin{equation}
{\bf P}
=
{\bf P}_0+\delta{\bf P}({\bf r},t),
\qquad
{\bf Q}
=
{\bf Q}_0+\delta{\bf Q}({\bf r},t),
\end{equation}
where the fluctuations are resolved into their normal modes as $\delta{\bf P}
=
\sum_{\bf q}
\delta{\bf P}_{\bf q}
e^{i{\bf q}\cdot{\bf r}-i\omega_P({\bf q})t}$ and 
$\delta{\bf Q}
=
\sum_{\bf q}
\delta{\bf Q}_{\bf q}
e^{i{\bf q}\cdot{\bf r}-i\omega_Q({\bf q})t}$. Here, $\omega_P({\bf q})$ and $\omega_Q({\bf q})$ denote dispersions of the polar and AFD modes. The corresponding static free-energy  is obtained by averaging
over the temporal fluctuations, 
$F=
\lim_{\mathcal{T}\rightarrow\infty}
\frac{1}{\mathcal{T}}
\int_{-\mathcal{T}/2}^{\mathcal{T}/2}
dt\,
\mathcal{H}(t)$. Retaining terms up to quadratic order in the fluctuations, the free
energy is decomposed as
\begin{equation}
F({\bf P},{\bf Q})
=\int{dr^3}
f_{\rm OP}({\bf P}_0,{\bf Q}_0)
+
F_{\rm FL}(\delta{\bf P},\delta{\bf Q}),
\end{equation}
where the uniform order-parameter contribution is
\begin{equation}
f_{\rm OP}({\bf P}_0,{\bf Q}_0)
=
\frac{a_P(T)}{2}P_0^2
+
\frac{u_P}{4}P_0^4
+
\frac{a_Q(T)}{2}Q_0^2
+
\frac{u_Q}{4}Q_0^4,
\end{equation}
and the quadratic fluctuation contribution  takes the form
\begin{eqnarray}
&&F_{\rm FL}(\delta{\bf P},\delta{\bf Q})
={}
\frac{1}{2}
\sum_{\bf q}
\left[
A_P({\bf q})
|\delta{\bf P}_{\bf q}|^2
+
A_Q({\bf q})
|\delta{\bf Q}_{\bf q}|^2
\right]
\nonumber\\
&&+
\sum_{\bf q}\left\{\lambda^{\rm eff}_{0}({\bf q})
\delta{\bf P}_{\bf q}\cdot
\left[{\bf Q}_0\times\left(i{\bf q}\times\delta{\bf Q}_{-{\bf q}}
\right)\right]+h.c.\right\}\nonumber\\
&&+
\sum_{\bf q}\left\{\lambda^{\rm eff}_{1}({\bf q})
\delta{\bf Q}_{\bf q}\cdot
\left[{\bf P}_0\times\left(i{\bf q}\times\delta{\bf P}_{-{\bf q}}
\right)\right]+h.c.\right\}.\label{FLEQ}
\end{eqnarray}
Thus, in the presence of a uniform AFD order parameter ${\bf Q}_0$ or polar one  ${\bf P}_0$, the original three-phonon interaction generates an effective bilinear coupling between polar and AFD fluctuations, which takes the form of a flexoelectric-like gradient coupling~\cite{axe1970anomalous,zubko2007strain,yudin2013fundamentals,guzman2023lamellar,morozovska2012interfacial,schiaffino2017macroscopic}. The fluctuation stiffnesses are
\begin{eqnarray}
A_P({\bf q})
&=&
a_P(T)
+
\frac{5}{3}u_PP_0^2
+
G_Pq^2
=
m_P\omega_P^2({\bf q}),
\\
A_Q({\bf q})
&=&
a_Q(T)
+
\frac{5}{3}u_QQ_0^2
+
G_Qq^2
=
m_Q\omega_Q^2({\bf q}).
\end{eqnarray}
The effective coupling strength $\lambda_{\rm eff}({\bf q})$ in the static sector is determined by the temporal
overlap between the polar and AFD fluctuations. For modes with a
finite lifetime, the sharp resonance condition is broadened and  the effective coupling entering the static fluctuation sector takes the form 
\begin{equation}
\lambda^{\rm eff}_{0,1}({\bf q})
=
\frac{\lambda_{0,1}}{\pi}
\frac{\Gamma^2}
{
[\omega_P({\bf q})-\omega_Q({\bf q})]^2+\Gamma^2
},\end{equation} 
where $\Gamma$ denotes the linewidth of the resonant mode coupling.

The equilibrium values $P_0$ and $Q_0$ are obtained by minimizing the uniform free energy, 
$\partial_{P_0}f_{\rm OP}=0$ and 
$\partial_{Q_0}f_{\rm OP}=0$.
 Substituting equilibrium values into $f_{\rm FL}$ determines the fluctuation spectrum and configurations around the uniform state. The wave vector at which the lowest hybridized mode reaches its minimum determines the characteristic spatial structure of the fluctuations. A minimum at ${\bf q}=0$ favors spatially uniform fluctuations, whereas a minimum at a finite wave vector $|{\bf q}|=q_0$ signals an instability toward spatially modulated polar textures~\cite{teubner1987origin} with a characteristic length scale
 $\xi_{\rm PNR}\sim{2\pi}/{q_0}$.

\begin{figure}[h]
    \centering
    \includegraphics[width=\linewidth]{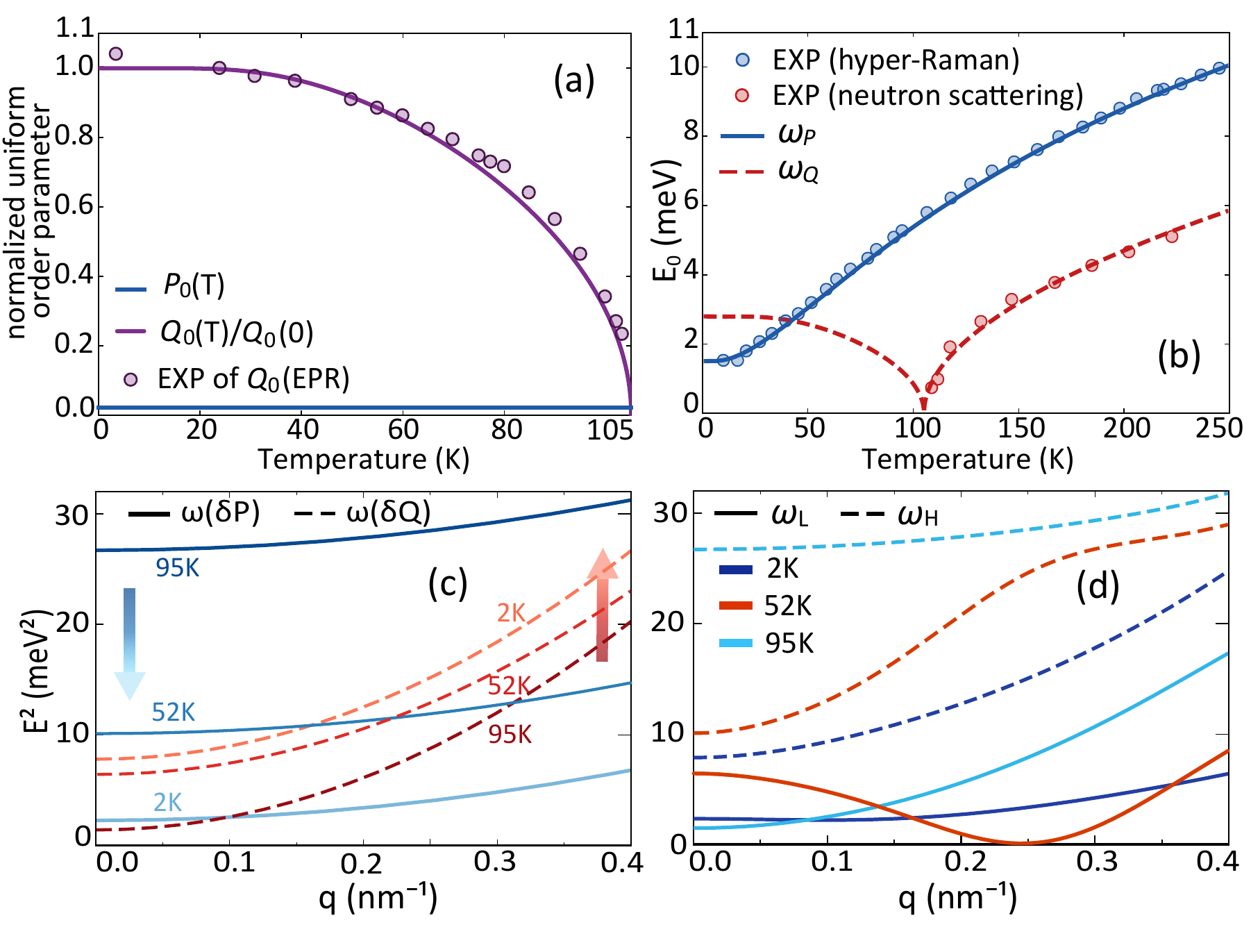}
    \caption{
    (a) Polar order parameter $P_0(T)$ and normalized AFD order parameter $Q_0(T)/Q_0(0)$, as well as  experimental octahedral rotation angles measured by electron paramagnetic resonance (EPR)~\cite{muller1968characteristic}.
    (b) Zone-center excitation energies of the polar (blue) and AFD (red) modes, compared with experimental data from hyper-Raman scattering~\cite{vogt1995refined} and neutron scattering~\cite{fauque2022mesoscopic}.
    (c) and (d) Bare and hybridized collective-mode dispersions, respectively, at 2, 52, and 95~K.}
    \label{fig:energy}
\end{figure}

{\sl Results.---}The model parameters and the temperature dependences of $\alpha_P(T)$ and $\alpha_Q(T)$ used in the calculations are summarized in Table~S1 of the Supplemental Material. The temperature-dependent quadratic coefficient $a_Q(T)$ changes sign at the AFD transition temperature $T_{\rm AFD}\simeq105$~K, with $a_Q(T)>0$ above $T_{\rm AFD}$ and $a_Q(T)<0$ below it. Consequently, a finite equilibrium AFD order parameter $Q_0(T)$ develops below $T_{\rm AFD}$, as shown in Fig.~\ref{fig:energy}(a). The polar coefficient $a_P(T)$ remains positive over the entire temperature range, yielding $P_0(T)\equiv0$ [Fig.~\ref{fig:energy}(a)]. In particular, upon cooling, $a_P(T)$ decreases and approaches a small but finite positive value as $T\rightarrow0$, characteristic of quantum-paraelectric ground state.

The Lifshitz coupling vanishes at ${\bf q}=0$ and therefore does not modify the zone-center excitation energies of the bare AFD and polar modes or alter the formation of the uniform order parameters. The temperature evolution of the corresponding zone-center excitation energies is shown in Fig.~\ref{fig:energy}(b). Upon cooling from high temperatures, the AFD mode progressively softens and becomes gapless at $T_{\rm AFD}\simeq105$~K, signaling the onset of long-range AFD order. Below $T_{\rm AFD}$, the development of the finite order parameter $Q_0(T)$ increases the curvature of the AFD potential, causing the AFD mode to harden again upon further cooling. In contrast, the polar mode continuously softens upon cooling while remaining gapped over the entire temperature range. Below approximately $20$~K, its excitation energy gradually saturates as thermal fluctuations are suppressed, approaching a small but finite value in the $T\rightarrow0$ limit, characteristic of the quantum-paraelectric ground state.

As seen in Fig.~\ref{fig:energy}(b), below $T_{\rm AFD}$ the two modes exhibit opposite temperature dependences~\cite{scott1997interpretation,courtens1997optical}: while the AFD mode hardens upon cooling, the polar mode continues to soften, causing their excitation energies to approach and eventually cross within an intermediate temperature window. Consequently, as shown in Fig.~\ref{fig:energy}(c), the crossing between the bare dispersions $\omega_P(q)$ and $\omega_Q(q)$ evolves strongly with temperature. At relatively high temperatures, for example $95$~K, the two modes intersect only at large $q$. Upon cooling, the crossing continuously shifts toward smaller $q$, reaching the long-wavelength regime around $45$-$59$~K. Below $\sim40$~K [Figs.~\ref{fig:energy}(b), (c)], the two bare dispersions become separated and no longer intersect, causing the resonance condition to disappear.

We then diagonalize the quadratic fluctuation contribution $f_{\rm FL}$ to obtain the hybridized collective-mode dispersions. Figure~\ref{fig:energy}(d) shows the resulting spectra at three representative temperatures, $95$~K, $52$~K, and $2$~K. At $95$~K, since the bare modes intersect only at much larger $q$, the hybridization predominantly affects short-wavelength fluctuations with a spatial scale too small to produce readily observable polar nanoregions. Upon cooling, the crossing shifts toward smaller wave vectors and the mode hybridization becomes most pronounced at intermediate temperatures. At $52$~K, the resonantly enhanced coupling strongly repels the two branches and pushes the lower hybridized branch downward. As a result, the low-energy branch develops a pronounced nonmonotonic minimum at a finite wave vector $q^*=0.25~{\rm nm}^{-1}$, with a minimum excitation energy of $0.037$~meV.  
 Upon further cooling, however, the two bare modes move away from resonance, and their crossing eventually disappears below approximately $40$~K. Consequently, the hybridization is strongly suppressed, the finite-$q$ minimum disappears, and the lower branch recovers a conventional monotonic dispersion, as illustrated at $2$~K.

\begin{figure}[h]
    \centering
    \includegraphics[width=\linewidth]{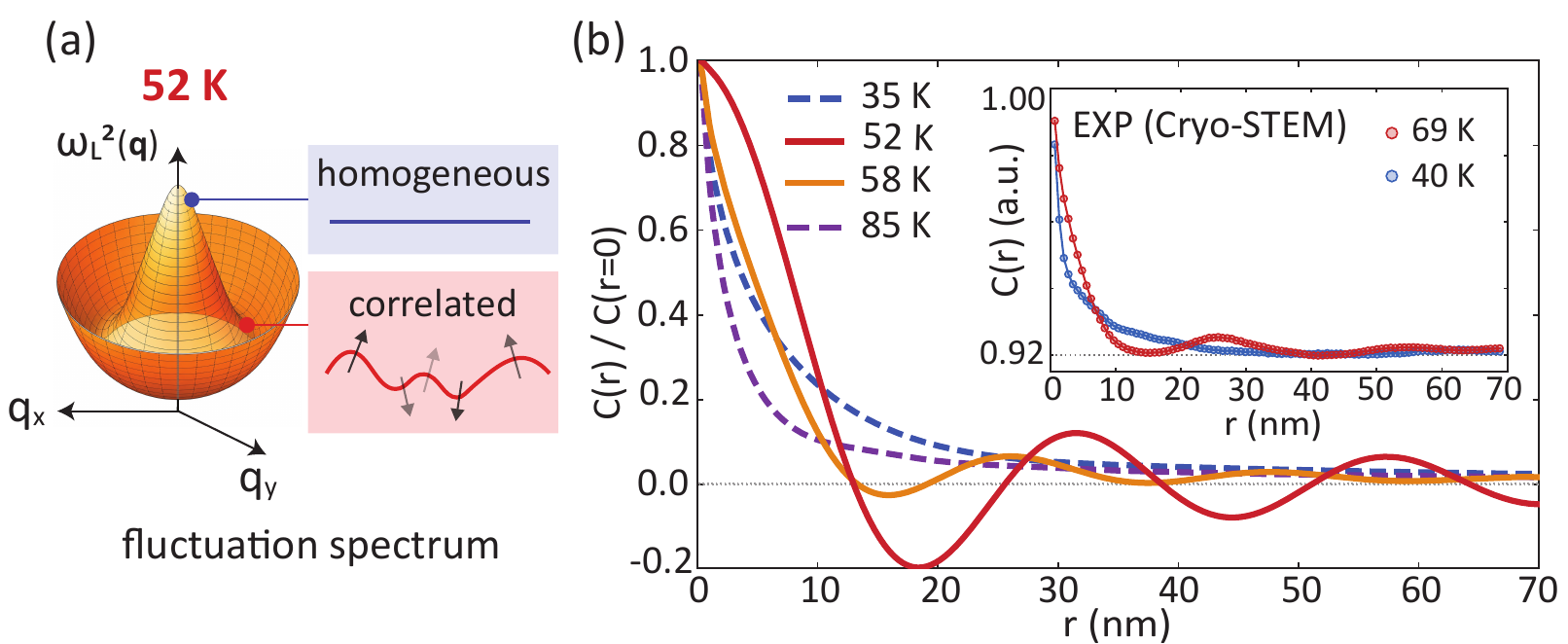}
    \caption{
    (a) Schematic $q_z=0$ section of the lower hybridized-mode dispersion at 52~K, showing a nearly degenerate ring of low-energy modes at finite $|{\bf q}|=q^*$. The finite-$q$ minimum favors spatially modulated polar correlations over a homogeneous configuration.
    (b) Calculated normalized real-space polar correlation function at 35, 52, 58, and 85~K. The oscillatory correlations are strongest near the mode resonance at 60~K and weaken away from the resonance temperature. Inset: experimentally measured polar correlations obtained by cryogenic scanning transmission electron microscopy (Cryo-STEM)~\cite{zhang2026imaging}.
    }
    \label{fig:correlation}
\end{figure}

The near softening of an entire shell of modes with $|{\bf q}|=q^*$ in the intermediate temperature window signals a Brazovskii-type finite-$q$ instability, as illustrated in Fig.~\ref{fig:correlation}(a), in sharp contrast to a conventional ferroelectric instability centered at ${\bf q}=0$. The finite-$q$ softening is directly reflected in the real-space polar correlations function $C({\bf r},T)
=\sum_{\bf q}
e^{i{\bf q}\cdot{\bf r}}
\left\langle
P({\bf q})P^*({\bf q})
\right\rangle_T$. 
As shown in Fig.~\ref{fig:correlation}(b), at both high and low temperatures, where no pronounced finite-$q$ softening occurs in the long-wavelength regime, $C(r)$ decays monotonically with distance, indicating the absence of a characteristic modulation length. In contrast, at $52$~K, the enhanced spectral weight around $q^*$ gives rise to pronounced oscillatory spatial correlations, revealing the emergence of finite-range modulated polar correlations. The selected wave vector corresponds to a real-space modulation period $\ell^*=2\pi/q^*\simeq25~{\rm nm}$, in good agreement with the characteristic periodicity extracted from the experimental polarization correlation function shown in the inset~\cite{zhang2026imaging}.

\begin{figure}[h]
    \centering
    \includegraphics[width=\linewidth]{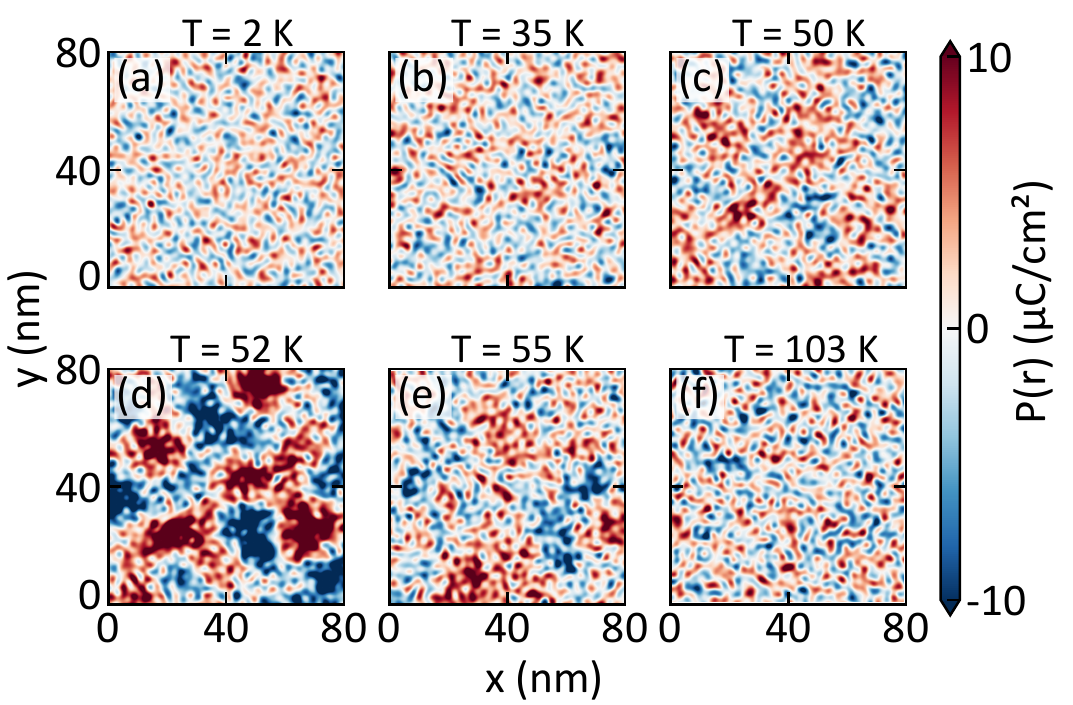}
    \caption{
    Real-space polarization textures $P({\bf r})$ at (a)-(f) 2, 35, 50, 52, 55, and 103~K. Spatially correlated polar regions become most pronounced near the mode resonance at 52~K.
    }
    \label{fig:Pr}
\end{figure}

To visualize polar correlations and the formation of PNRs, we generate real-space polarization configurations $P({\bf r})$ from the fluctuation spectrum $G_{\rm eff}(q,T)=2\langle P({\bf q})P^*({\bf q})\rangle_T$ at different temperatures. Considering the spectrum structure of the finite-$q$ instability illustrated in Fig.~\ref{fig:correlation}(a), where the low-energy fluctuations are distributed over an approximately degenerate spherical shell in 3D momentum space, using the random-phase spectral representation method~\cite{roberts1997statistical,berk1987scattering}, we discretize the radial spectrum into wavenumber shell and evaluate the angular contribution by Monte Carlo sampling over uniformly distributed propagation directions~\cite{makse1996method} (see Sec.~SV). 
 
The resulting representative polarization configurations are shown in Fig.~\ref{fig:Pr}. First,  the coupling between the polar and AFD fluctuations in Eq.~(\ref{FLEQ}) relies on the formation of a finite AFD order parameter, and hence,  can only emerge below the AFD transition temperature, $T_{\rm AFD}\simeq 105~$K. Upon cooling from $105$~K, the initially short-range and spatially disordered polarization fluctuations progressively develop spatial correlations as the system approaches the mode-resonance window. At 45-59~K, nanoscale correlated regions become clearly visible. In particular, near resonance at $52$~K, the finite-$q$ spectral weight is maximally enhanced, causing neighboring polarization fluctuations to organize into extended regions with the same polarization sign over tens of nanometers. The characteristic modulation visible in real space is consistent with the wavelength $2\pi/q^*\simeq25$~nm selected by the soft finite-$q$ mode. Remarkably, these nanoscale polar structures do not continue to strengthen upon further cooling below 52~K, and the extended correlated regions begin to weaken as the two collective modes move away from resonance. At $35$~K, the large-scale spatial organization is largely lost, and upon further cooling the polarization configuration again becomes dominated by short-range, spatially disordered fluctuations. The real-space evolution therefore directly reproduces the reentrant behavior observed experimentally: PNRs emerge upon cooling below $T_{\rm AFD}$, become most pronounced within the intermediate-temperature resonance window, and disappear again toward the low-temperature quantum-paraelectric state. This behavior is fully consistent with the experimental observations~\cite{zhang2026imaging}.

{\sl Discussion.---}The reentrant-like evolution identified here is not accompanied by a thermodynamic phase transition, since the interaction is not sufficiently strong to soften the lower hybridized mode  into an imaginary-frequency instability.
 It remains associated with a positive excitation energy, similar to the conventional optical phonon mode. Thus, the system does not spontaneously select a particular ordering wave vector and develop a modulated stripe phase (polar density waves~\cite{orenstein2025observation,Yang2026PDW}).  The emergence and disappearance of PNRs represent thermal crossovers in the spatial organization of fluctuations, with no additional symmetry breaking or uniform ferroelectric order involved. The nonmonotonicity upon cooling resides in how fluctuations are distributed in momentum space and organized in real space. Near resonance, mode hybridization concentrates low-energy fluctuations around a characteristic finite wave vector, allowing otherwise disordered local fluctuations to develop correlations over tens of nanometers and hence nanoscale polar textures while the uniform polarization remains zero.  Further cooling detunes the polar and AFD modes, removes this finite-$q$ concentration of spectral weight, and dissolves the associated nanoscale correlations.  

These results demonstrate that the magnitude of fluctuations and their spatial organization are fundamentally distinct quantities, and the degree of local spatial organization need not evolve monotonically with the overall fluctuation strength. A system can develop stronger local correlations while its total fluctuations are being suppressed, and subsequently lose those correlations upon further cooling, entirely through a redistribution of fluctuation spectral weight among collective modes. Reentrant nanoscale order can therefore emerge without a reentrant thermodynamic phase transition and disappear without any conventional disorder.  Accordingly, the reentrant weakening behavior of PNR requires no anomalous accumulation of fluctuations toward zero temperature, as suggested by experimental interpretation~\cite{zhang2026imaging}. The overall fluctuation strength continues to decrease upon cooling, consistent with the thermodynamic reduction of entropy. In this sense, resonance between competing instabilities provides a unique route by which cooling can first organize fluctuations into finite-length-scale structures and subsequently dissolve them as the underlying excitation spectrum evolves.

The present mechanism also leads to an unusual prediction when SrTiO$_3$ is driven slightly across its ferroelectric quantum critical point. Taking a weak strain of $0.02\%$, slightly above the reported quantum critical value of approximately $0.019\%$~\cite{li2025classical,wang2026beyond}, the ground state becomes ferroelectric, while the intermediate-temperature resonant enhancement of finite-$q$ polar correlations remains operative. After a full simulation under this weak strain, we find a striking nonmonotonic evolution upon heating, as shown in Fig.~\ref{fig:strain}. Upon heating, the low-temperature ferroelectric phase first gives way to a paraelectric regime, followed by the emergence of PNRs at intermediate temperatures and their eventual disappearance in the high-temperature paraelectric state. This ferroelectric-paraelectric-PNR-paraelectric sequence is particularly unusual because  ferroelectric order and nanoscale polar correlations emerge in distinct temperature windows under a single monotonic variation of temperature, giving rise to a reentrant-like evolution of polar order without requiring a second ferroelectric phase transition. In other words, the pronounced PNR regime is not continuously connected to the ferroelectric phase in SrTiO$_3$, but emerges as an isolated temperature window of local polar organization embedded within the paraelectric state. This suggests that competing instabilities can
create an isolated ``island'' of spatial order inside an otherwise disordered
regime, a route to reentrant-like behavior fundamentally different from the
usual competition between thermodynamic phases.

\begin{widetext}
  \begin{center}
\begin{figure}[h]
    \centering
\includegraphics[width=17.5cm]{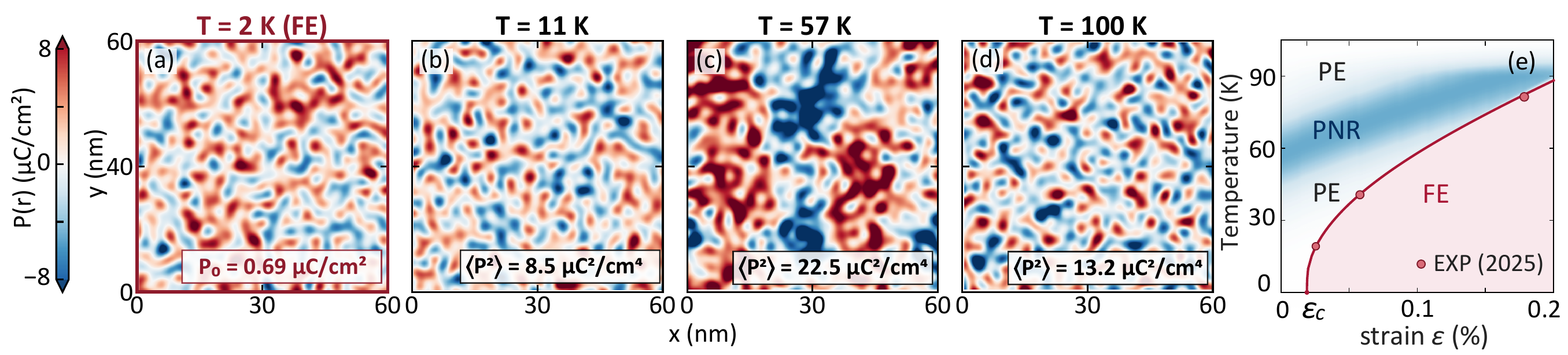}
   \caption{
Real-space \(P(\mathbf{r})\) at (a)-(d) 2, 11, 57, and
100~K with external strain 0.02\% showing the reentrant-like evolution of polar ordering. Panel (e) shows a temperature-strain phase diagram with ferroelectric phase transition. The experimental data of $T_c$ comes from Ref.~\cite{li2025classical}.
}
\label{fig:strain}
\end{figure}
\end{center}
\end{widetext}

{\it Acknowledgments.---}F.Y. and Y.J.S. performed the theoretical modeling and numerical simulations, respectively, and contributed equally to this work.  Y.J.S. was supported by the National Science Foundation  under Grants No. DMR-2133373. F.Y. and L.Q.C. acknowledge support from the US Department of Energy, Office of Science, Basic Energy Sciences, under Award
Number DE-SC0020145 as part of Computational Materials
Sciences Program. F.Y. and L.Q.C. also appreciate the generous support from the Donald W. Hamer Foundation through a Hamer Professorship at Penn State.

\begin{acknowledgments}

\end{acknowledgments}

\bibliography{sample}

\end{document}


\title{Polar nanoregions and reentrant-like ferroelectric behavior in  SrTiO$_3$ \\ (Supplementary Materials)}   

\author{Yuan-Jie Sun}
\affiliation{Department of Materials Science and Engineering and Materials Research Institute, The Pennsylvania State University, University Park, PA 16802, USA}

\author{Fei Yang}
\email{fzy5099@psu.edu}

\affiliation{Department of Materials Science and Engineering and Materials Research Institute, The Pennsylvania State University, University Park, PA 16802, USA}

\author{Long-Qing Chen}
\email{lqc3@psu.edu}

\affiliation{Department of Materials Science and Engineering and Materials Research Institute, The Pennsylvania State University, University Park, PA 16802, USA}

\maketitle 
\tableofcontents

\section{Polarization and antiferrodistortive (AFD) distortion}

\subsection{Polarization}

In displacive ferroelectricity, a polar lattice distortion originates from the transverse optical (TO) soft mode at
the center of the Brillouin zone ($\Gamma$ point), and  the ferroelectric instability can be understood as the condensation of
this zone-center polar soft mode. As the TO mode softens, its harmonic
restoring force decreases. When the squared frequency of the
$\Gamma$-point mode becomes negative, 
$\omega_{\rm TO}^2({\bf q}=0)<0$, the centrosymmetric configuration becomes unstable and the equilibrium
position shifts to a state with a finite static amplitude of the
zone-center soft mode. Denoting the corresponding soft-mode coordinate
by $\phi_{\rm sp}$, the equilibrium displacement amplitude can be written
as
\begin{equation}
\xi_0
=
\big|
\big\langle
\phi_{\rm sp}({\bf q}=0)
\big\rangle
\big|.
\label{eq:soft-mode-condensation}
\end{equation}
Thus, condensation of the ${\bf q}=0$ polar mode corresponds
microscopically to a coherent ionic displacement extending throughout
the crystal. For a polar mode with ionic displacement direction ${\bf e}_{u}$, the
resulting macroscopic polarization is related to the soft-mode
displacement by
\begin{equation}
{\bf P}_0
=
\frac{z^*{\bf e}_{u}}{\Omega_{\rm cell}}
\xi_0,
\label{eq:polarization-displacement}
\end{equation}
where $z^*$ is the mode effective charge and $\Omega_{\rm cell}$ is the
unit-cell volume. This relation provides the microscopic connection between the ionic
soft-mode coordinates and the continuum polarization field used in the
effective theory. The same correspondence is employed in
first-principles effective-Hamiltonian approaches, where local soft-mode
amplitudes are introduced as the fundamental lattice degrees of freedom
and converted into polarization through their associated mode effective
charges.

Consequently, the polarization field can be decomposed as
\begin{equation}
{\bf P}({\bf r},t)
=
{\bf P}_0(T)+\delta{\bf P}({\bf r},t),
\qquad
\langle\delta{\bf P}({\bf r},t)\rangle=0,
\end{equation}
where ${\bf P}_0(T)=\langle{\bf P}({\bf r},t)\rangle$ represents the
condensed ${\bf q}=0$ component of the polar mode. A finite ${\bf P}_0$ corresponds to long-range ferroelectric order. In quantum paraelectrics such as SrTiO$_3$, the $\Gamma$-point TO soft
mode strongly softens upon cooling but remains uncondensed. Consequently,
the equilibrium polarization remains ${\bf P}_0=0$, while finite
fluctuations of the polar soft mode persist. 

\subsection{Antiferrodistortive (AFD) distortion}

The tetragonal antiferrodistortive (AFD) phase of SrTiO$_3$ is
characterized by rotations of neighboring TiO$_6$ octahedra about
a common axis in opposite senses~\cite{sai2000first}. This alternating
pattern can be described by a slowly varying axial-vector order
parameter $\mathbf Q$. To distinguish the continuum amplitude from the
local octahedral rotation, the octahedra are labeled by their Ti
positions in the parent cubic reference lattice,
$\mathbf R_\ell=a_{\rm lat}(n_x,n_y,n_z)$, where $a_{\rm lat}$ is
the reference lattice parameter and $n_x,n_y,n_z$ are integers.
The alternating rotation pattern is associated with the $R$-point
wave vector
\begin{equation}
\mathbf q_R=\frac{\pi}{a_{\rm lat}}(1,1,1),
\end{equation}
with the corresponding phase factor 
$s_\ell=e^{i\mathbf q_R\cdot\mathbf R_\ell}
=(-1)^{n_x+n_y+n_z}$. Thus, $s_\ell$ changes sign between nearest-neighbor Ti sites.
The physical rotation vector of octahedron $\ell$ can then be written as
\begin{equation}
\boldsymbol\theta_\ell
=s_\ell\theta(\mathbf R_\ell)
\widehat{\mathbf n}(\mathbf R_\ell),
\label{eq:local-rotation-vector}
\end{equation}
where $\theta\geq0$ is the rotation angle in radians and the oriented
unit vector $\widehat{\mathbf n}$ specifies the rotation axis and sense
on the reference sublattice $s_\ell=+1$. Following the displacement normalization of Ref.~\cite{sai2000first},
the continuum AFD order parameter is defined as
\begin{equation}
\mathbf Q(\mathbf r)
=
\frac{a_{\rm lat}}{2}\sin\theta(\mathbf r)
\widehat{\mathbf n}(\mathbf r)
=
\frac{a_{\rm lat}}{2}\theta(\mathbf r)
\widehat{\mathbf n}(\mathbf r)
+\mathcal O(a_{\rm lat}\theta^3).
\label{eq:Q-angle-vector}
\end{equation}
A spatially uniform $\mathbf Q$ represents a uniform
amplitude of the underlying staggered $R$-point rotation pattern.

\section{Hamiltonian and free energy}

The effective potential Hamiltonian of coupled polar and AFD fields,
$\mathbf P$ and $\mathbf Q$ can be written as 
\begin{equation}
\mathcal{H}
=\int\,d^3r
\left[\mathcal{V}_P({\bf P})
+\mathcal{V}_Q({\bf Q})
+\mathcal{V}_{\rm int}({\bf P},{\bf Q})
\right],
\end{equation}
where the potential densities  associated with the polarization and AFD distortions as well as their coupling are given by 
\begin{eqnarray}
\mathcal V_P({\bf P})&=&
\frac{a_P(T)}{2}|\mathbf P|^2+
\frac{u_P}{4}|\mathbf P|^4+
\frac{G_P}{2}(\partial_iP_j)(\partial_iP_j),
\\
\mathcal V_Q({\bf Q})&=&
\frac{a_Q(T)}{2}|\mathbf Q|^2+
\frac{u_Q}{4}|\mathbf Q|^4+
\frac{G_Q}{2}(\partial_iQ_j)(\partial_iQ_j).
\end{eqnarray}
The polar and AFD lattice modes can be described by orthogonal normal coordinates of the harmonic lattice Hamiltonian of the centrosymmetric parent phase. We therefore focus on their leading symmetry-allowed anharmonic coupling. At cubic order, this interaction takes the form of a symmetry-allowed three-phonon coupling between the polar and AFD branches, representing the conventional lowest-order anharmonic interaction between coupled lattice modes in solid-state theory. Specifically, since $\mathbf P$ is a polar vector, whereas
$\mathbf Q$, describing the AFD octahedral rotation, is an axial
vector, a local cubic coupling of the form $PQQ$ is forbidden by
inversion symmetry of the centrosymmetric parent structure. The leading symmetry-allowed cubic interaction therefore involves
a spatial gradient and has the generic form
$P_i Q_j \partial_k Q_l$. An analogous argument applies to the
$QPP$ coupling. Moreover, in the present work, we focus on the transverse polar and AFD
phonon branches, 
$\nabla\cdot\delta\mathbf P=0$, and 
$\nabla\cdot\delta\mathbf Q=0$, for which the relevant spatial variations are represented by the
rotational components $\nabla\times\mathbf P$ and
$\nabla\times\mathbf Q$. Contracting the vector indices to construct
scalar invariants gives the symmetry-allowed three-phonon
anharmonic interactions
\begin{equation}
\mathcal V_{\mathrm{int}}({\bf P},{\bf Q})
=
\lambda_0\mathbf P\cdot
\left[\mathbf Q\times(\nabla\times\mathbf Q)\right]
+
\lambda_1\mathbf Q\cdot
\left[\mathbf P\times(\nabla\times\mathbf P)\right].
\label{eq:three-phonon-interaction}
\end{equation}
The first and second terms correspond, respectively, to gradient-assisted
$PQQ$ and $QPP$ three-phonon vertices. In momentum space, the spatial
gradient produces a factor $i\mathbf q$, so that these interactions
vanish in the uniform limit and couple finite-momentum transverse
lattice fluctuations.  The absence of a clear anomaly in the experimentally observed polar phonon energy and measured dielectric constant at $T_{\mathrm{AFD}}=105~$K in SrTiO$_3$ suggests that the quartic coupling $|\mathbf{P}|^2|\mathbf{Q}|^2$ contributes only weakly. We therefore neglect this term and other higher-order couplings in the model. 

We expand the order parameters about their uniform configurations,
\begin{equation}
{\bf P}
=
{\bf P}_0+\delta{\bf P}({\bf r},t),
\qquad
{\bf Q}
=
{\bf Q}_0+\delta{\bf Q}({\bf r},t),
\end{equation}
and retaining terms up to quadratic
order in the fluctuations, the coupling term becomes 
\begin{align}
\mathcal V_{\mathrm{int}}^{(2)}
={}&
\lambda_0\delta\mathbf P\cdot
\left[\mathbf Q_0\times(\nabla\times\delta\mathbf Q)\right]
+
\lambda_0\mathbf P_0\cdot
\left[\delta\mathbf Q\times(\nabla\times\delta\mathbf Q)\right]+
\lambda_1\delta\mathbf Q\cdot
\left[\mathbf P_0\times(\nabla\times\delta\mathbf P)\right]
+
\lambda_1\mathbf Q_0\cdot
\left[\delta\mathbf P\times(\nabla\times\delta\mathbf P)\right].
\label{eq:SI-expanded-interaction}
\end{align}
For transverse fluctuations,
$\nabla\cdot\delta\mathbf P=\nabla\cdot\delta\mathbf Q=0$,
periodic boundaries imply
\begin{equation}
\int d^3r\,
\delta\mathbf P\times(\nabla\times\delta\mathbf P)=0,
\qquad
\int d^3r\,
\delta\mathbf Q\times(\nabla\times\delta\mathbf Q)=0.
\end{equation}
The second and fourth terms in
Eq.~\eqref{eq:SI-expanded-interaction} consequently give no
bulk contribution. The first and third terms survive,
producing bilinear gradient couplings whose amplitudes are
proportional to $\mathbf Q_0$ and $\mathbf P_0$, respectively.

To obtain the corresponding static free energy, the fluctuations are
resolved in the orthogonal normal-mode basis of the uncoupled polar and
AFD modes,
\begin{equation}
\delta{\bf P}
=
\sum_{\bf q}
\delta{\bf P}_{\bf q}
e^{i{\bf q}\cdot{\bf r}-i\omega_P({\bf q})t}, \quad \delta{\bf Q}
=
\sum_{\bf q}
\delta{\bf Q}_{\bf q}
e^{i{\bf q}\cdot{\bf r}-i\omega_Q({\bf q})t},
\end{equation}
where $\omega_P({\bf q})$ and $\omega_Q({\bf q})$ denote the bare
dispersions of the polar and AFD modes, respectively. The orthogonality
of the spatial Fourier modes ensures momentum conservation, while the
subsequent temporal averaging determines the frequency overlap between
the two collective modes. Then, by averaging
over the temporal fluctuations, 
$F=
\lim_{\mathcal{T}\rightarrow\infty}
\frac{1}{\mathcal{T}}
\int_{-\mathcal{T}/2}^{\mathcal{T}/2}
dt\,
\mathcal{H}(t)$, the free
energy is decomposed as
\begin{equation}
F
=\int{dr^3}
f_{\rm OP}({\bf P}_0,{\bf Q}_0)
+
F_{\rm FL}(\delta{\bf P},\delta{\bf Q}),
\end{equation} 
where the uniform order-parameter contribution is
\begin{equation}
f_{\rm OP}
=
\frac{a_P(T)}{2}P_0^2
+
\frac{u_P}{4}P_0^4
+
\frac{a_Q(T)}{2}Q_0^2
+
\frac{u_Q}{4}Q_0^4,
\end{equation}
and the quadratic fluctuation contribution  takes the form
\begin{eqnarray}
\!\!F_{\rm FL}
\!=\!\frac{1}{2}
\sum_{\bf q}
\left[
A_P({\bf q})
|\delta{\bf P}_{\bf q}|^2
\!+\!
A_Q({\bf q})
|\delta{\bf Q}_{\bf q}|^2
\right]\!+\!
\sum_{\bf q}\left\{\lambda^{\rm eff}_{0}({\bf q})
\delta{\bf P}_{\bf q}\!\cdot\!
\left[{\bf Q}_0\!\times\!\left(i{\bf q}\!\times\!\delta{\bf Q}_{-{\bf q}}
\right)\right]\!+\!\lambda^{\rm eff}_{1}({\bf q})
\delta{\bf Q}_{\bf q}\!\cdot\!
\left[{\bf P}_0\!\times\!\left(i{\bf q}\!\times\!\delta{\bf P}_{-{\bf q}}
\right)\right]\!+\!h.c.\right\}.\label{FLEQ}
\end{eqnarray}
Here, the fluctuation stiffnesses, after performing an isotropic average over the three-component fluctuations, are given by
\begin{eqnarray}
A_P({\bf q})=
a_P(T)
+
\frac{5}{3}u_PP_0^2
+
G_Pq^2,\quad
A_Q({\bf q})=
a_Q(T)
+
\frac{5}{3}u_QQ_0^2
+
G_Qq^2,
\end{eqnarray}
which determines the bare dispersions of the polar and AFD modes self-consistently, i.e., $A_P({\bf q})=
m_P\omega_P^2({\bf q})$ and $A_Q({\bf q})=
m_Q\omega_Q^2({\bf q})$. The factor $5/3$ appearing in both $A_P({\bf q})$ and
$A_Q({\bf q})$ originates from the quadratic expansion of the
quartic self-interaction of a three-component vector order parameter
about its uniform equilibrium state. Specifically, using
the AFD field $\mathbf Q$ as example, one has
\begin{align}
|\mathbf Q|^2
=Q_0^2+2\mathbf Q_0\cdot\delta\mathbf Q+\delta Q^2,\quad
|\mathbf Q|^4=Q_0^4
+4Q_0^2(\mathbf Q_0\cdot\delta\mathbf Q)
+2Q_0^2\delta Q^2
+4(\mathbf Q_0\cdot\delta\mathbf Q)^2
+\mathcal O(\delta Q^3).
\label{eq:Q-fourth-expansion}
\end{align}
Substituting these expressions into 
$f_Q=
\frac{a_Q}{2}|\mathbf Q|^2
+\frac{u_Q}{4}|\mathbf Q|^4$ and grouping terms according to the order of the fluctuations gives
\begin{equation}
\begin{aligned}
f_Q={}&
\underbrace{
\frac{a_Q}{2}Q_0^2+\frac{u_Q}{4}Q_0^4
}_{f_Q^{(0)}}
+
\underbrace{
(a_Q+u_QQ_0^2)
\mathbf Q_0\cdot\delta\mathbf Q
}_{f_Q^{(1)}}+
\underbrace{
\frac12(a_Q+u_QQ_0^2)\delta Q^2
+
u_Q(\mathbf Q_0\cdot\delta\mathbf Q)^2
}_{f_Q^{(2)}}
+\mathcal O(\delta Q^3).
\end{aligned}
\label{eq:afd-expansion}
\end{equation}
The uniform equilibrium condition ensures that the linear fluctuation term $f_Q^{(1)}$ vanishes. For isotropic fluctuations of the three-component AFD field,
the angular average satisfies 
$\left\langle
(\mathbf Q_0\cdot\delta\mathbf Q)^2
\right\rangle
=
\frac{1}{3}Q_0^2
\left\langle\delta Q^2\right\rangle$. The quadratic contribution therefore becomes $\langle f_Q^{(2)}\rangle
=\frac12(
a_Q+\frac{5}{3}u_QQ_0^2)\langle\delta Q^2\rangle$, yielding the fluctuation stiffness $A_{Q}({\bf q})$ above. \\

The corresponding static contribution of the interaction terms is obtained
by averaging over the temporal fluctuations,
\begin{align}
F_{\rm int}
=
\lim_{\mathcal T\rightarrow\infty}
\frac{1}{\mathcal T}
\int_{-\mathcal T/2}^{\mathcal T/2}dt
\sum_{\bf q}
\Bigg\{
\lambda_0
{\bf P}_{\bf q}\cdot
\left[
{\bf Q}_0\times
\left(
-i{\bf q}\times{\bf Q}_{-{\bf q}}
\right)
\right]
e^{-i\Delta\omega_{\bf q}t}
+
\lambda_1
{\bf Q}_{\bf q}\cdot
\left[
{\bf P}_0\times
\left(
-i{\bf q}\times{\bf P}_{-{\bf q}}
\right)
\right]
e^{+i\Delta\omega_{\bf q}t}
+
{\rm h.c.}
\Bigg\},
\label{eq:temporal-average-interaction}
\end{align}
where 
$\Delta\omega_{\bf q}
=
\omega_P({\bf q})-\omega_Q({\bf q})$. Then, only frequency-matched modes
satisfying 
$\omega_P({\bf q})=\omega_Q({\bf q})$ give a nonvanishing static contribution, whereas off-resonant
contributions average to zero. For modes with finite lifetimes, the frequency-matching
condition is broadened. Introducing an exponentially decaying temporal
coherence $e^{-\Gamma |t|}$ gives the normalized temporal overlap
\begin{align}
\mathcal R({\bf q})
&=
\frac{
\displaystyle
\int_{-\infty}^{\infty}dt\,
e^{-\Gamma|t|}
e^{\pm i\Delta\omega_{\bf q}t}
}{
\displaystyle
\int_{-\infty}^{\infty}dt\,
e^{-\Gamma|t|}
}=
\frac{\Gamma^2}
{
[\omega_P({\bf q})-\omega_Q({\bf q})]^2+\Gamma^2
}.
\label{eq:temporal-overlap}
\end{align}
Consequently, the effective coupling strengths are
\begin{equation}
\lambda_{0,1}^{\rm eff}({\bf q})
=
\lambda_{0,1}
\frac{\Gamma^2}
{
[\omega_P({\bf q})-\omega_Q({\bf q})]^2+\Gamma^2
}.
\label{eq:effective-coupling}
\end{equation}

\section{Eigenenergies of coupled polar-AFD }
\label{sec:coupled-eigenenergies}

\subsection{Transverse fluctuation sector}

For the transverse polar and AFD branches, the fluctuation amplitudes satisfy 
$\mathbf q\cdot\delta\mathbf P_{\mathbf q}=0$ and $\mathbf q\cdot\delta\mathbf Q_{\mathbf q}=0$. Then, the gradient couplings can be simplified as
\begin{equation}
\begin{aligned}
\delta\mathbf P_{\mathbf q}\cdot
\left[
\mathbf Q_0\times
(i\mathbf q\times\delta\mathbf Q_{\mathbf q}^{*})
\right]=
-i(\mathbf Q_0\cdot\mathbf q)
\delta\mathbf P_{\mathbf q}\cdot
\delta\mathbf Q_{\mathbf q}^{*},\quad
\delta\mathbf Q_{\mathbf q}\cdot
\left[
\mathbf P_0\times
(i\mathbf q\times\delta\mathbf P_{\mathbf q}^{*})
\right]=
-i(\mathbf P_0\cdot\mathbf q)
\delta\mathbf Q_{\mathbf q}\cdot
\delta\mathbf P_{\mathbf q}^{*}.
\end{aligned}
\label{eq:eigen-transverse-identities}
\end{equation}

Substituting these relations into $f_{\mathrm{FL}}$ and combining the
Hermitian-conjugate terms gives
\begin{equation}
\begin{aligned}
F_{\rm FL}=\frac12\sum_{\mathbf q}
\left[
A_P|\delta\mathbf P_{\mathbf q}|^2
+A_Q|\delta\mathbf Q_{\mathbf q}|^2
\right]+i\sum_{\mathbf q}
\left[
(\lambda_0^{\rm eff}\mathbf Q_0
-\lambda_1^{\rm eff}\mathbf P_0)\cdot\mathbf q
\right]
\left(
\delta\mathbf P_{\mathbf q}^{*}\cdot\delta\mathbf Q_{\mathbf q}
-\delta\mathbf Q_{\mathbf q}^{*}\cdot\delta\mathbf P_{\mathbf q}
\right).
\end{aligned}
\label{eq:eigen-transverse-energy}
\end{equation}

\subsection{Eigenenergies}

The dynamics of the coupled polar and AFD fluctuations is governed by
the restoring forces derived from the quadratic fluctuation free energy
$F_{\rm FL}$. For conservative lattice dynamics, the equations of motion
are written as
\begin{equation}
m_P\partial_t^2\delta P_{\mathbf q}
=
-\frac{\delta F_{\rm FL}}
{\delta \delta P_{\mathbf q}^{*}},
\qquad
m_Q\partial_t^2\delta Q_{\mathbf q}
=
-\frac{\delta F_{\rm FL}}
{\delta \delta Q_{\mathbf q}^{*}},
\label{eq:fluctuation-equations-general}
\end{equation}
yielding 
\begin{align}
m_P\partial_t^2\delta P_{\mathbf q}
={}&
-A_P(\mathbf q)\delta P_{\mathbf q}
-2i
\left[
\left(
\lambda_0^{\rm eff}\mathbf Q_0
-\lambda_1^{\rm eff}\mathbf P_0
\right)\cdot\mathbf q
\right]
\delta Q_{\mathbf q},
\label{eq:eigen-motion-P}
\\
m_Q\partial_t^2\delta Q_{\mathbf q}
={}&
-A_Q(\mathbf q)\delta Q_{\mathbf q}
+2i
\left[
\left(
\lambda_0^{\rm eff}\mathbf Q_0
-\lambda_1^{\rm eff}\mathbf P_0
\right)\cdot\mathbf q
\right]
\delta P_{\mathbf q}.
\label{eq:eigen-motion-Q}
\end{align}
Here, $m_P$ and $m_Q$ are the inertia densities associated with the
polar and AFD mode coordinates, respectively.
The equations can be written in the generalized eigenvalue form
\begin{equation}
\begin{pmatrix}
A_P &
2i[(\lambda_0^{\rm eff}\mathbf Q_0
-\lambda_1^{\rm eff}\mathbf P_0)\cdot\mathbf q]
\\
-2i[(\lambda_0^{\rm eff}\mathbf Q_0
-\lambda_1^{\rm eff}\mathbf P_0)\cdot\mathbf q] &
A_Q
\end{pmatrix}
\begin{pmatrix}
\delta P_{\mathbf q}\\
\delta Q_{\mathbf q}
\end{pmatrix}
=
\omega^2
\begin{pmatrix}
m_P&0\\
0&m_Q
\end{pmatrix}
\begin{pmatrix}
\delta P_{\mathbf q}\\
\delta Q_{\mathbf q}
\end{pmatrix}.
\label{eq:eigen-generalized-problem}
\end{equation}

The eigenfrequencies are therefore determined by
\begin{equation}
(A_P-m_P\omega^2)(A_Q-m_Q\omega^2)
-4\left[
(\lambda_0^{\rm eff}\mathbf Q_0
-\lambda_1^{\rm eff}\mathbf P_0)\cdot\mathbf q
\right]^2=0.
\label{eq:eigen-characteristic}
\end{equation}

For SrTiO$_3$, the equilibrium polarization vanishes,
$\mathbf P_0=\mathbf 0$. For a single tetragonal AFD domain, we define
\begin{equation}
\Lambda(\mathbf q,T)
=
\frac{2\hbar^2}{\sqrt{m_Pm_Q}}
\lambda_0^{\rm eff}(q,T)
\left[{\mathbf q}\cdot\mathbf Q_0(T)
\right].
\label{eq:reduced-coupling}
\end{equation}
The corresponding eigenenergies are
\begin{equation}
\begin{aligned}
E_\pm^2(\mathbf q,T)
=
\frac12
\Bigg[\frac{\hbar^2A_P(q,T)}{m_P}
+\frac{\hbar^2A_Q(q,T)}{m_Q}\pm
\sqrt{
\left[
\frac{\hbar^2A_P(q,T)}{m_P}
-\frac{\hbar^2A_Q(q,T)}{m_Q}
\right]^2
+4\Lambda^2(\mathbf q,T)
}
\Bigg].
\end{aligned}
\label{eq:coupled-eigenenergies}
\end{equation}

In bulk SrTiO$_3$, the tetragonal AFD order can occur in three
symmetry-equivalent orientational domains with
\begin{equation}
\mathbf Q_0^{(x)}=Q_0\widehat{\mathbf x},
\qquad
\mathbf Q_0^{(y)}=Q_0\widehat{\mathbf y},
\qquad
\mathbf Q_0^{(z)}=Q_0\widehat{\mathbf z}.
\end{equation}
For an equally populated multidomain sample, as commonly observed
experimentally in bulk SrTiO$_3$, the domain average of the coupling
entering the eigenenergies is therefore
\begin{equation}
\begin{aligned}
\left\langle
\left({\mathbf q}\cdot\mathbf Q_0
\right)^2
\right\rangle_{\rm dom}=
\frac{Q_0^2}{3}
\left(q_x^2
+q_y^2
+q_z^2
\right)=
\frac{Q_0^2q^2}{3}.
\end{aligned}
\label{eq:afd-domain-average}
\end{equation}
Accordingly, for quantities averaged over equivalent tetragonal domains,
the squared coupling may be replaced by
\begin{equation}
\Lambda^2(\mathbf q,T)
\rightarrow
{\Lambda}^2(q,T)
=
\frac{4\hbar^4q^2}{3m_Pm_Q}
\left[\lambda_0^{\rm eff}(q,T)\right]^2
Q_0^2(T).
\label{eq:domain-averaged-coupling}
\end{equation}
The resulting domain-averaged eigenenergies are isotropic and depend only
on the magnitude $q=|\mathbf q|$.

\section{Correlation function}
\label{sec:correlation}

Equation~\eqref{eq:coupled-eigenenergies} gives the two eigenvalues of each
transverse block.  To obtain the polarization fluctuations, we next need the
corresponding eigenvectors. 
Let $\psi_+$ and $\psi_-$ denote the independent unit-inertia normal
coordinates.  Transformation gives
\begin{equation}
\begin{pmatrix}\mathcal P\\\mathcal Q\end{pmatrix}=U\begin{pmatrix}\psi_+\\\psi_-\end{pmatrix},\qquad U=\begin{pmatrix}\dfrac{1}{\sqrt{1+\alpha_+^2}}&\dfrac{1}{\sqrt{1+\alpha_-^2}}\\[8pt]\dfrac{-\mathrm i\alpha_+}{\sqrt{1+\alpha_+^2}}&\dfrac{-\mathrm i\alpha_-}{\sqrt{1+\alpha_-^2}}\end{pmatrix}.
\label{eq:corr-normal-transformation}
\end{equation}
where
\begin{equation}
\alpha_\pm(\mathbf q,T)=\frac{{\hbar^2A_Q(q,T)}/{m_Q}-{\hbar^2A_P(q,T)}/{m_P}\pm\sqrt{\left[{\hbar^2A_Q(q,T)}/{m_Q}-{\hbar^2A_P(q,T)}/{m_P}\right]^2+4\Lambda^2(\mathbf q,T)}}{2q\Lambda(\mathbf q,T)}.
\label{eq:alpha-pm}
\end{equation}

In particular, the physical polarization is
\begin{equation}
P=\frac{1}{\sqrt{m_P}}\mathcal{P}=\frac{1}{\sqrt{m_P}}
\left(
\frac{\psi_+}{\sqrt{1+\alpha_+^2}}
+\frac{\psi_-}{\sqrt{1+\alpha_-^2}}
\right).
\label{eq:polarization-normal-modes}
\end{equation}

Equation~\eqref{eq:polarization-normal-modes} gives the polarization
correlation in terms of the two hybridized normal modes:
\begin{equation}
\begin{aligned}
\left\langle P P^* \right\rangle
=
\frac{1}{m_P}
\Bigg[
&\frac{\left\langle |\psi_+|^2 \right\rangle}{1+\alpha_+^2}
+
\frac{\left\langle \psi_+\psi_-^* \right\rangle}
{\sqrt{(1+\alpha_+^2)(1+\alpha_-^2)}}+
\frac{\left\langle \psi_-\psi_+^* \right\rangle}
{\sqrt{(1+\alpha_+^2)(1+\alpha_-^2)}}
+
\frac{\left\langle |\psi_-|^2 \right\rangle}{1+\alpha_-^2}
\Bigg].
\end{aligned}
\label{eq:polar-correlation-full}
\end{equation}

Since $\psi_+$ and $\psi_-$ are independent normal modes of the
quadratic Hamiltonian, the cross correlations vanish,
\begin{equation}
\left\langle \psi_+\psi_-^* \right\rangle
=
\left\langle \psi_-\psi_+^* \right\rangle
=0.
\end{equation}
The equilibrium fluctuation amplitude of each normal mode, according to fluctuation-dissipation theorem, is
\begin{equation}
\left\langle |\psi_\pm|^2 \right\rangle
=
\frac{\hbar^2}{2E_\pm}
\coth\left(
\frac{E_\pm}{2k_{\mathrm B}T}
\right).
\label{eq:normal-mode-fluctuation}
\end{equation}
Therefore, the polarization fluctuation spectrum of one transverse
polarization branch is
\begin{equation}
\left\langle P P^* \right\rangle
=
\frac{\hbar^2}{2m_P}
\left[
\frac{1}{E_+(1+\alpha_+^2)}
\coth\left(\frac{E_+}{2k_{\mathrm B}T}\right)
+
\frac{1}{E_-(1+\alpha_-^2)}
\coth\left(\frac{E_-}{2k_{\mathrm B}T}\right)
\right].
\label{eq:polar-fluctuation-spectrum}
\end{equation}

Including the two transverse polarization branches gives  $G^{\mathrm{eff}}(\mathbf q,T)
=
2\left\langle P P^* \right\rangle$. The transverse polarization correlation function used in the main text is
\begin{equation}
C(\mathbf r,T)
=
\left\langle
\mathbf P_T(\mathbf r)\cdot
\mathbf P_T(\mathbf 0)
\right\rangle
=
\int\frac{\mathrm d^3q}{(2\pi)^3}
e^{i\mathbf q\cdot\mathbf r}
G^{\mathrm{eff}}(\mathbf q,T).
\label{eq:C-three-dimensional}
\end{equation}

\section{Real-space polarization configuration}
\label{sec:random-field}

To visualize the spatial structure associated with the polarization
correlations, we generate real-space polarization configurations whose
ensemble-averaged spectrum reproduces the calculated transverse
polarization spectrum ${G}^{\rm eff}(q,T)$.
We employ the random-phase spectral representation
method~\cite{shinozuka1991simulation,shinozuka1996simulation}.  Specifically, the momentum interval is divided into spherical shells with centers
$q_m$ and widths $\Delta q_m$. For each shell, $N_{\Omega}$ independent
propagation directions are sampled uniformly over the unit sphere.
A scalar realization of the polarization fluctuation is constructed as
\begin{equation}
P(\mathbf r,T)
=
\sum_{m=1}^{N_q}
\sum_{n=1}^{N_{\Omega}}
A_m(T)
\cos\!\left(
q_m\hat{\mathbf n}_{mn}\cdot\mathbf r+\phi_{mn}
\right),
\label{eq:random-mode-sum}
\end{equation}
where the phases are independent random variables, $\phi_{mn}\sim\mathcal U(0,2\pi)$. The propagation directions are sampled uniformly over the unit sphere
according to
\begin{equation}
u_{mn}\sim\mathcal U(-1,1),
\qquad
\chi_{mn}\sim\mathcal U(0,2\pi),
\qquad
\hat{\mathbf n}_{mn}
=
\begin{pmatrix}
\sqrt{1-u_{mn}^2}\cos\chi_{mn},
\sqrt{1-u_{mn}^2}\sin\chi_{mn},
u_{mn}
\end{pmatrix}.
\label{eq:uniform-sphere}
\end{equation}
Since the phases are statistically independent, the contribution of shell $m$ to the local variance is
therefore
\begin{equation}
\left\langle P_m^2(\mathbf r)\right\rangle_{\phi}
=
\frac{N_{\Omega}A_m^2}{2}.
\label{eq:shell-variance}
\end{equation}
On the other hand, 
the equal-position polarization correlation is 
$\left\langle P^2(\mathbf r)\right\rangle
=
\int\frac{\mathrm d^3q}{(2\pi)^3}{G}^{\rm eff}(q,T)$, and for an isotropic spectrum, the contribution from a spherical shell of
radius $q_m$ and width $\Delta q_m$ is
\begin{equation}
\left\langle P_m^2(\mathbf r)\right\rangle
=
\frac{q_m^2\Delta q_m}{2\pi^2}{G}^{\rm eff}(q_m,T).
\label{eq:spectral-shell-variance}
\end{equation}
Equating this expression with
Eq.~\eqref{eq:shell-variance} determines the spectral amplitude,
\begin{equation}
A_m(T)
=
\Big[
\frac{q_m^2\Delta q_m}{\pi^2N_{\Omega}}{G}^{\rm eff}(q_m,T)
\Big]^{1/2}.
\label{eq:three-dimensional-mode-amplitude}
\end{equation}
Thus, the ensemble average of the random-phase configurations reproduces
the polarization fluctuation spectrum obtained from the coupled
polar-AFD modes.

\section{Model parameters and simulation details}

Using the equilibrium conditions for the uniform polar and AFD order
parameters, i.e., $P_0^2(T)=-{\alpha_P(T)}/{u_P}$ when $\alpha_P(T)<0$, with $P_0=0$ for $\alpha_P(T)>0$, and
$Q_0^2(T)=-{a_Q(T)}/{u_Q}$,
when $T<T_{\mathrm{AFD}}$, with $Q_0=0$ for $T>T_{\mathrm{AFD}}$, the corresponding fluctuation stiffnesses are
\begin{equation}
A_P(q,T)
=
\begin{cases}
\alpha_P(T)+G_Pq^2,
& \alpha_P(T)>0, \\[6pt]
\dfrac{2}{3}|\alpha_P(T)|+G_Pq^2,
& \alpha_P(T)<0,
\end{cases}\qquad
A_Q(q,T)
=
\begin{cases}
a_Q(T)+G_Qq^2,
& T>T_{\mathrm{AFD}}, \\[6pt]
\dfrac{2}{3}|a_Q(T)|+G_Qq^2,
& T<T_{\mathrm{AFD}}.
\end{cases}
\label{eq:afd-stiffness}
\end{equation}
The temperature dependence of the polar stiffness is determined {\sl 
self-consistently}. Following Ref.~\cite{yang2025hotphonon},
\begin{equation}
\alpha_P(T)
=
\alpha_P(0)
+
\frac{5}{3}u_PC(T),
\label{eq:alphaP-temperature}
\end{equation}
where the thermal correction to the harmonic coefficient is
\begin{equation}
C(T)
=
\frac{\hbar}{m_P}
\sum_{\mathbf q}
\left[
\frac{1}{\omega_q(T)}
\coth\left(
\frac{\hbar\omega_q(T)}{2k_{\mathrm B}T}
\right)
-
\frac{1}{\omega_q(0)}
\right].
\label{eq:polar-fluctuation-correction}
\end{equation}
The bare polar-mode frequency entering Eq.~\eqref{eq:polar-fluctuation-correction}
is determined by the corresponding fluctuation stiffness,
\begin{equation}
\omega_q^2(T)
=
\frac{A_P(q,T)}{m_P},
\label{eq:polar-dispersion}
\end{equation}
where $A_P(q,T)$ is given by Eq.~\eqref{eq:afd-stiffness}.
Therefore, Eqs.~\eqref{eq:alphaP-temperature}-
\eqref{eq:polar-dispersion} form a closed set of self-consistent equations:
the fluctuation correction $C(T)$ renormalizes $\alpha_P(T)$, which
determines $A_P(q,T)$ and hence bare $\omega_q(T)$, while $\omega_q(T)$ in
turn determines $C(T)$. These equations are iterated until convergence.

For the AFD sector, the temperature-dependent quadratic coefficient
$a_Q(T)$ is phenomenologically parametrized by the quantum-saturation form
\begin{equation}
a_Q(T)
=\eta_Q\left[
\coth\left(\frac{T_s}{T}\right)
-
\coth\left(\frac{T_s}{T_{\rm AFD}}\right)
\right],
\label{eq:aQ-quantum-saturated}
\end{equation}
following Refs.~\cite{salje1991order,hayward1999cubic}. Here $A>0$ sets the stiffness scale, $T_s$ controls the
low-temperature saturation, and $T_{\rm AFD}$ is the AFD transition
temperature. This parametrization satisfies
$a_Q(T_{\rm AFD})=0$ and approaches a finite negative value as
$T\rightarrow0$.

For the real-space visualization, the field is evaluated on a
$256\times256$ grid spanning an $80~\mathrm{nm}$ square, corresponding
to the $z=0$ plane of the three-dimensional random-mode construction.
The radial momentum grid contains 480 shells from
$10^{-3}~\text{\AA}^{-1}$ to $q_c=0.20~\text{\AA}^{-1}$, with
$N_{\Omega}=64$ propagation directions sampled for each shell.

\begin{table*}[h!]
\centering
\caption{
Parameters of the coupled polar-AFD model for bulk SrTiO$_3$. For the AFD sector, $T_s$ is fitted to the normalized temperature dependence of the
octahedral rotation angle~\cite{muller1968characteristic}. The stiffness amplitude $\eta_Q$ and the effective inertia $m_Q$
is determined by the AFD soft-mode
energies above $T_{\rm AFD}$~\cite{fauque2022mesoscopic}. The parameter $q_{\max}=0.58\,\text{\AA}^{-1}$ is the upper
radial wavevector cutoff of correlation integral Eq.~\eqref{eq:C-three-dimensional}.
The applied strain in the model $\epsilon=0.02\%$ is close to the critical strain $\epsilon_c=0.019\%$~\cite{li2025classical}, thus the ferroelectric transition temperature $T_c=10.1$~K is well below AFD transition temperature $T_{c(\mathrm{\epsilon=0.02\%})}\ll T_{\mathrm{AFD}}=105$~K. We therefore assume
$|\lambda_1^{\rm eff}P_0|\ll|\lambda_0^{\rm eff}Q_0|$ and neglect the
$\lambda_1^{\rm eff}\mathbf P_0$ contribution to the polar-AFD
coupling. }
\centering
\begin{minipage}{0.95\textwidth}
\centering
\label{tab:model-parameters}
\small
\setlength{\tabcolsep}{3pt}
\renewcommand{\arraystretch}{1.35}
\begin{ruledtabular}
\begin{tabular}{@{}cccccc@{}}

Polar
& $m_P$ ($\mathrm{meV\,ps^2}\,\text{\AA}/e^2$)
& $G_P$ ($\mathrm{meV}\,\text{\AA}^3/e^2$)
& $u_P$ ($\mathrm{meV}\,\text{\AA}^5/e^4$)
& $q_{\max}$ ($\text{\AA}^{-1}$)
& $\alpha_{P}(0)$ (meV$\cdot${\AA}/e$^2$) \\ sector 
& $2.13299$~\cite{khalsa2012theory}
& $1.39946\times10^4$~\cite{khalsa2012theory,yang2026self}
& $3.14195\times10^6$~\cite{yang2026self}
& $0.58$
& $5.061\times10^{-5}/\varepsilon_0$ \\[5pt]
\hline

AFD
& $m_Q$ ($\mathrm{meV\,ps^2}/\text{\AA}^5$)
& $G_Q$ ($\mathrm{meV}/\text{\AA}^3$)
& $\eta_Q$ ($\mathrm{meV}/\text{\AA}^5$)
& $T_s$ (K)
& $T_{\mathrm{AFD}}$ (K) \\
sector & $0.0187119$
& $4.43771\times10^2$~\cite{morozovska2012interfacial}
& $0.845687$
& $77.1804$
& $105$~\cite{hayward1999cubic} \\[5pt]
\hline

Coupling
& $\lambda_0$ ($\mathrm{meV}/(e\,\text{\AA}^2)$)
& $\Gamma$ ($\mathrm{ps}^{-1}$)
& & & \\
sector & $8.55772\times10^3$
& $1.36734$
& & & \\[5pt]
\hline

Strain
& $\epsilon_c$ (\%)
& $\epsilon$ (\%)
& & & \\
sector & $0.019$~\cite{li2025classical}
& $0.020$
& & & \\

\end{tabular}
\end{ruledtabular}
\end{minipage}
\end{table*}

To account for the effect of strain, we  introduce a
strain dependence of the zero-temperature polar stiffness as
\begin{equation}
\alpha_P(0,\epsilon)
=
\alpha_P(0,0)
\left(
1-\frac{\epsilon}{\epsilon_c}
\right),
\label{eq:strain-polar-stiffness}
\end{equation}
where $\epsilon$ is the applied strain and $\epsilon_c$ denotes the
critical strain at which the zero-temperature polar stiffness vanishes,
corresponding to a strain-tuned ferroelectric quantum critical point.
Experimentally, this quantum critical point in SrTiO$_3$ occurs at a
strain of approximately $0.019\%$. We therefore take $\epsilon_c=0.019\%$ in the present calculations.
Accordingly, in the presence of strain, $\alpha_P(0)$ in the above
self-consistent equations is replaced by $\alpha_P(0,\epsilon)$.
Then, strain acts as an external tuning parameter for the polar
instability, while the finite-temperature polar stiffness and soft-mode
dispersion are determined by the same self-consistent fluctuation
renormalization described above.

Throughout this work, we focus on the low-lying transverse polar fluctuations,
which satisfy 
$\mathbf q\cdot\delta\mathbf P_{\mathbf q}=0$. Such transverse fluctuations do not generate a longitudinal bound-charge
density. Consequently, the nonanalytic long-range Coulomb interaction, which
provides an additional restoring force for the longitudinal polar mode
and gives rise to the LO-TO splitting, does not enter the transverse
sector considered in our work.

\bibliography{sample.bib}